\documentclass[preprint,1p,11pt]{StyleFiles/ISAS_IR}

\pdfoutput=1

\input{StyleFiles/Packages.sty}
\input{StyleFiles/Abbreviations.sty}
\input{StyleFiles/Defs.sty}
\input{StyleFiles/Format.sty}

\begin{document}

\begin{frontmatter}

\title{FLUX: Progressive State Estimation Based on\\
Zakai-type Distributed Ordinary Differential Equations}

\author{Uwe~D.~Hanebeck}
\ead{Uwe.Hanebeck@kit.edu}

\address{%
	Intelligent Sensor-Actuator-Systems Laboratory (ISAS)\\
	Institute for Anthropomatics and Robotics\\
	Karlsruhe Institute of Technology (KIT), Germany
	}%

\begin{abstract}
	%
%
We propose a homotopy continuation method called FLUX for approximating complicated probability density functions.
%
%
It is based on progressive processing for smoothly morphing a given density into the desired one.
%
%
Distributed ordinary differential equations (DODEs) with an artificial time $\gamma \in [0,1]$ are derived for describing the evolution from the initial density to the desired final density.
%
%
For a finite-dimensional parametrization, the DODEs are converted to a system of ordinary differential equations (SODEs), which are solved for $\gamma \in [0,1]$ and return the desired result for $\gamma=1$.
%
%
This includes parametric representations such as Gaussians or Gaussian mixtures and nonparametric setups such as sample sets.
%
%
In the latter case, we obtain a particle flow between the two densities along the artificial time.

%
%
FLUX is applied to state estimation in stochastic nonlinear dynamic systems
%
%
by gradual inclusion of measurement information.
%
%
The proposed approximation method 
(1) is fast, 
(2) can be applied to arbitrary nonlinear systems and is not limited to additive noise,
(3) allows for target densities that are only known at certain points,
(4) does not require optimization,
(5) does not require the solution of partial differential equations, and
(6) works with standard procedures for solving SODEs.
%
%
This manuscript is limited to the one-dimensional case and a fixed number of parameters during the progression. Future extensions will include consideration of higher dimensions and on the fly adaption of the number of parameters.
\end{abstract}

\end{frontmatter}

%
%

%
%

\section{Introduction} \label{Sec_Introduction}

%
%
We consider the approximation of complicated probability density functions (pdfs) by densities with convenient finite-dimensional representations. 
This includes parametric density approximation such as Gaussian densities or Gaussian mixture densities and nonparametric representations such as sample sets.
%
%
Three problem classes are investigated:
(1)~The fundamental problem of approximating a given probability density function that may only be given at certain points.
(2)~The problem of approximating posterior densities in estimating a hidden state based on prior densities and given measurements.
(3)~The problem of propagating a state estimate through a discrete-time stochastic nonlinear dynamic system.

%
%
Optimal approximation of complicated probability density functions is a hard problem and typically involves the minimization of a suitable distance measure between the original density and its approximation. 
The minimization problem is in general nonlinear and nonconvex, which precludes efficient solution procedures and calls for iterative optimization methods. 
Depending on the selected starting points, the run time varies and one might end up in local minima.

%
%
In this manuscript, our goal is to find a flow between a simple density with a given approximation and the target density. 
This flow is then used to move the parameters of the approximating density in such a way that they finally approximate the target density.

%
%

\subsection{State of the Art} \label{Sec_StateOfArt}

%
%

\subsubsection{Overview of Relevant Literature} \label{Sec_OverviewLiterature}

%
%
When using a particle representation for the posterior probability density functions in a Bayesian update step, it is well known that a narrow likelihood function corresponding to a precise measurement typically leads to particle degeneration \cite{daumParticleDegeneracyRoot2011}. 
This means that due to the downweighting with the likelihood function, only a few particles maintain nonzero weights so that all the other particles are wasted as they do not contribute to the density approximation.

%
%
Particle degeneration can be mitigated with a discrete set of so-called bridging densities that allow a gradual introduction of the measurement by employing them successively with intermediate resampling as opposed to the direct multiplication of a particle set with the original likelihood \cite{oudjaneProgressiveCorrectionRegularized2000}.
%
%
This strategy is used by annealed sampling methods \cite{nealAnnealedImportanceSampling2001,delmoralSequentialMonteCarlo2006,gallInteractingAnnealingParticle2007}.
They use a form of MCMC or sample movement to cope with the particle degeneracy. Particle-flow filters work in a fundamentally different fashion as they do not employ particle weighting at all.

%
%
The Daum-Huang filters \cite{daumNonlinearFiltersLoghomotopy2007,daumNonlinearFiltersParticle2009a,daumNonlinearFiltersParticle2009} use a homotopy between the logarithms of the unnormalized prior and posterior densities, which is used to calculate a particle flow. 
Different versions have been proposed, see \cite{daumExactParticleFlow2010,daumNumericalExperimentsCoulomb2011,daumParticleFlowInspired2013}.
The flow moves particles towards regions of a high posterior probability. Its calculation entails the solution of a partial differential equation, which is computationally complex.
Two types of Daum-Huang filters are compared with standard filter approaches in \cite{khanNonlinearNonGaussianState2014}.
To cope with the high computational complexity of particle flows, an approximate Gaussian flow is introduced in \cite{bunchApproximationsOptimalImportance2014}. Here, the particles are not used directly for representing the posterior density. Instead, they are used as an input to an importance sampler.

%
%
Related to particle-flow filters are the so-called feedback particle filters \cite{yangFeedbackParticleFilter2013} that apply a feedback mechanism in order to move the particles. 
According to \cite{berntorpFeedbackParticleFilter2018}, these filters are a generalization of
linear-regression filters such as the UKF \cite{julierNewMethodNonlinear2000} or the smart-sampling Kalman filter first proposed in \cite{Fusion13_Steinbring}, refined in \cite{JAIF14_Steinbring-S2KF}, and optimized by enforcing symmetric samples in \cite{JAIF16_Symmetric_S2KF_Steinbring}. 
The feedback gains of the feedback particle filters are obtained by minimizing a cost function based on the Kullback-Leibler divergence.
An evaluation of feedback particle filters is given in \cite{berntorpFeedbackParticleFilter2015}.

%
%

\subsubsection{Own Prior Work} \label{Sec_OwnPriorWork}

%
%
A novel estimation method based on progressively introducing measurement information in order to perform morphing between a given density and a desired density, in this case, the posterior resulting from a Bayesian update step, is introduced in \cite{SPIE03_HanebeckBriechle-ProgBayes,MFI03_Hanebeck}. 
A generalization to multi-dimensional states is given in \cite{CDC03_Feiermann-ProgBayes}. 
In \cite{hagmarOptimalParameterizationPosterior2011}, this method was generalized to different distance measures.
%
%
Further applications of progressive processing methods besides state estimation are in moment calculation \cite{SPL03_Rauh} and Gaussian mixture reduction \cite{Fusion08_Huber-PGMR}.

%
%
A progressive Bayesian procedure for Gaussian-assumed density filtering, where the measurement information is  gradually included into the given prior estimate is proposed in \cite{arXiv12_Hanebeck,Fusion12_Hanebeck}. 
The solution is a special case of the framework proposed in this manuscript. 
A flow for the parameters of the posterior Gaussian density is derived and compared with the state of the art.

%
%
A different approach based on a discrete set of bridging densities is pursued in \cite{Fusion11_Ruoff,Fusion16_Hanebeck}. 
Here, the likelihood function is decomposed into a product of wider likelihoods that are applied sequentially. 
After each application and a resulting slight reweighting of the particles, the particles are reapproximated with an equally weighted set. 
This translates the successive reweighting into a particle motion. 
For re-approximation, a special type of distance, a so-called permutation-invariant set-distance is employed \cite{AT15_Hanebeck} for comparing two particle sets. 

%
%
As the online minimization of the distance measure is rather costly, a simplified progression method is used in 
\cite{Fusion13_Hanebeck} in the context of Gaussian filters. It is called the Progressive Gaussian Filter (PGF).
Here, the re-approximation of weighted particles is performed by replacing them with a pre-calculated set of equally weighted particles corresponding to the same underlying Gaussian density.
The original PGF directly works with the measurement equation, no likelihood is required. A more efficient variant for the case that the likelihood function is explicitly available is given in \cite{Fusion14_Steinbring}.
A semi-analytic variant of the PGF for measurement equations with a conditionally linear part is derived in \cite{MFI16_Steinbring}.
The PGF can easily be parallelized, which is exploited for a GPU-based implementation in \cite{Fusion15_Steinbring}. 

%
%

\subsection{Contributions} \label{Sec_Contributions}

%
%
The contributions of this manuscript is a homotopy continuation method called FLUX for approximating complicated probability density functions.
%
%
FLUX is based on a progressive mechanism for smoothly morphing a given density $f_0$ into a target density $f_1$.
%
%
%
An approximation for $f_0$ is already known. The goal is to find an approximation of $f_1$.
%
%
The evolution of the densities from $f_0$ to $f_1$ is described by distributed ordinary differential equations (DODEs) with an artificial time $\gamma \in [0,1]$. 
These DODEs are expressed in terms of infinite-dimensional density representations that are impractical for computer implementation and can only be solved in special cases.
%
%
In the case of a finite-dimensional parametrization, the DODEs can be converted to a system of ordinary differential equations (SODEs), which are solved for $\gamma \in [0,1]$ and return the desired result for $\gamma=1$.
%
%
This includes parametric representations such as Gaussians or Gaussian mixtures and nonparametric setups such as sample sets.
%
%
In the latter case, we obtain a particle flow between the two densities along the artificial time.

%
%
FLUX is applied to various density approximation methods. This includes approximating the posterior densities in the measurement update step for state estimation in stochastic nonlinear dynamic systems
%
%
by gradual inclusion of measurement information.
%
%
The proposed particle flow method 
\begin{itemize}
\item is fast, 
\item can be applied to arbitrary nonlinear systems and is not limited to additive noise,
\item allows for target densities that are only known at certain points,
\item does not require optimization,
\item does not require the solution of partial differential equations, and
\item  works with standard procedures for solving SODEs.
\end{itemize}

%
%

\subsection{Structure of the Paper} \label{Sec_Structure}

This manuscript is structured as follows. 
In \Sec{Sec_ProbForm}, a rigorous formulation of the pursued approximation problem is given, which includes the most fundamental setup of morphing between two arbitrary densities and also its application to state estimation in stochastic nonlinear dynamic systems.
The corresponding distributed ordinary differential equations (DODEs) for describing the evolution of the densities in the infinite-dimensional case are derived in \Sec{Sec_Derivation_Zakai_DODE}. 
Exact solutions for special cases are given in \Sec{Sec_Exact_Soln}.
Finite-dimensional approximate solutions are derived for parametric density representations in \Sec{Sec_Parametric_Soln} and for nonparametric density representations in \Sec{Sec_Nonparametric_Soln}.
Conclusion are given in \Sec{Sec_Conclude} followed by an outlook to future work.

%
%

\section{Problem Formulation} \label{Sec_ProbForm}

%
%
The most fundamental problem we will consider is the approximation of a density $f_1(\vec{x})$ by some parametrized density $f_1(\vec{x}, \vec{\eta}_1)$ with parameter vector $\vec{\eta}_1$. 
We assume that a somehow related density $f_0(\vec{x})$ exists that is already approximated by a density $f_0(\vec{x}, \vec{\eta}_0)$.
%
%
Our goal is to find a way to transfer the given approximation $f_0(\vec{x}, \vec{\eta}_0)$ of $f_0(\vec{x})$ to $f_1(\vec{x})$ in order to find $\vec{\eta}_1$.

%
%
In the following, we will consider special cases of related densities $f_0(\vec{x})$ and $f_1(\vec{x})$ by estimating posterior densities in the context of stochastic nonlinear dynamic systems. 
We start with the measurement update or filter step that uses measurements to update the state estimate in \SubSec{Subsec_ProbForm_MeasurementUpdate}. 
In \SubSec{Subsec_ProbForm_TimeUpdate}, we consider the time update or prediction step, where the state estimate is propagated through a nonlinear function and corrupted by noise.

%
%

\subsection{Measurement Update} \label{Subsec_ProbForm_MeasurementUpdate}

%
%
We consider a discrete-time system with a hidden state $\vec{x}_k$ at time step $k$ that is observed via a measurement equation
\begin{equation} \label{Eq_Measurement_general}
    \vec{y}_k = \vec{h}_k(\vec{x}_k, \vec{v}_k)
\end{equation}
with measurement noise $\vec{v}_k$ and a nonlinear measurement function $\vec{h}_k(.,.)$. 
%
%
We assume that a probabilistic description of the measurement equation is given in form of a likelihood function $f_k^L(\vec{y}_k \,|\, \vec{x}_k)$.
%
%
We are given a prior density
\begin{equation*}
    f_k^p(\vec{x}_k) = f(\vec{x}_k \,|\, \vec{y}_{1:k-1})
\end{equation*}
that summarizes our knowledge about the state based on all past measurements 
\begin{equation*}
    \vec{y}_{1:k-1} = \begin{bmatrix} \vec{y}_1, \vec{y}_2, \ldots, \vec{y}_{k-2}, \vec{y}_{k-1} \end{bmatrix}^\top
\end{equation*}
up to the previous time step.
%
%
We are interested in the posterior density at time step $k$, which is defined as the density of the state $\vec{x}_k$ 
\begin{equation*}
    f_k^e(\vec{x}_k) = f(\vec{x}_k \,|\, \vec{y}_{1:k})
\end{equation*}
conditioned on all measurements up to the current time step 
\begin{equation*}
    \vec{y}_{1:k} = \begin{bmatrix} \vec{y}_1, \vec{y}_2, \ldots, \vec{y}_{k-1}, \vec{y}_k \end{bmatrix}^\top
    \enspace .
\end{equation*}
Via Bayes' law, the desired posterior can be expressed as the normalized product of the prior density before the measurement update and the likelihood describing the measurement equation as
\begin{equation} \label{Eq_Posterior_normalized}
    f_k^e(\vec{x}_k) = c_k \, f_k^L(\vec{y}_k \,|\, \vec{x}_k) f_k^p(\vec{x}_k) \enspace ,
\end{equation}
where $c_k$ is a normalization constant.

%
%

\subsubsection{Additive Noise} \label{SubSubSec_ProbForm_AdditiveNoise}

%
%
For additive measurement noise, the measurement equation can be written as
\begin{equation} \label{Eq_Measurement_additive}
    \vec{y}_k = \vec{h}_k(\vec{x}_k) + \vec{v}_k \enspace .
\end{equation}
The measurement noise $\vec{v}_k$ is assumed to be zero mean Gaussian with covariance matrix $\mat{C}_k^v$ or standard deviation $\sigma_k^v$ in the scalar case. 
For the scalar case, the likelihood function is then given as
\begin{equation*}
    f_k^L(y_k \,|\, x_k) 
    = \exp\left( -\frac{1}{2} \frac{(y_k-h_k(x_k))^2}{(\sigma_k^v)^2} \right)
\end{equation*}
and in general as
\begin{equation*}
    f_k^L(\vec{y}_k \,|\, \vec{x}_k) 
    = \exp\left( -\frac{1}{2} \big( \vec{y}_k-\vec{h}_k(\vec{x}_k) \big)^\top
    \big( \mat{C}_k^v \big)^{-1} \big( \vec{y}_k-\vec{h}_k(\vec{x}_k) \big) \right)
    \enspace ,
\end{equation*}
where all multiplicative constants are omitted.

%
%

\subsubsection{Non-additive Noise}

%
%
In general, the structure of the measurement equation is not as simple as in the previous subsection. 
Typically, the noise does not enter additively. We will take a look at two examples: Purely multiplicative noise in \Ex{Ex_likelihoodMult} and combined multiplicative and additive noise in \Ex{Ex_likelihoodMultAdd}.

%
%
\begin{example}[Multiplicative Noise] \label{Ex_likelihoodMult}
As an example, we first consider the case of multiplicative noise for scalar states, i.e., the measurement equation is given by
\begin{equation*}
    y_k = x_k \, v_k \enspace ,
\end{equation*}
with $v_k \sim f_k^v(v_k)$. 
The conditional density of the measurement $y_k$ given the state $x_k$ and the noise $v_k$ is given by 
\begin{equation*}
    f(y_k \,|\, x_k, v_k) = \delta(y_k - x_k \, v_k) \enspace .
\end{equation*}
The conditional density $f(y_k, v_k \,|\, x_k)$ is given by
\begin{equation*}
    f(y_k, v_k \,|\, x_k) = \delta(y_k - x_k \, v_k) f_k^v(v_k) \enspace ,
\end{equation*}
so that we obtain
\begin{equation*}
    f(y_k \,|\, x_k) = \int_\NewR \delta(y_k - x_k \, v_k) f_k^v(v) \, \mathrm{d} v_k \enspace ,
\end{equation*}
which is equivalent to
\begin{equation*}
    f(y_k \,|\, x_k) = \int_\NewR \frac{1}{ |x_k| } 
    \delta\left( v_k - \frac{y_k}{x_k} \right) f_k^v(v_k) \, \mathrm{d} v \enspace .
\end{equation*}
We finally obtain
\begin{equation} \label{Eq_likelihoodMult}
    f(y_k \,|\, x_k) = \frac{1}{ |x_k| } f_k^v\left( \frac{y_k}{x_k} \right) \enspace ,
\end{equation}
which is the likelihood function $f_k^L(x_k)$ for a given $y_k$.
\end{example}

%
%
We will take a look at a second example that generalizes the one on multiplicative noise. 
The insights will be used when deriving a progressive update for the multiplicative noise case.

%
%
\begin{example}[Multiplicative and Additive Noise] \label{Ex_likelihoodMultAdd}
We now consider the case of a measurement corrupted by multiplicative and additive noise as
\begin{equation*}
    y_k = x_k \, v_k + w_k \enspace ,
\end{equation*}
with $v_k,w_k \sim f_k^{vw}(v_k,w_k)$. The conditional density of the measurement $y_k$ given the state $x_k$ and the noise terms $v_k$ and $w_k$ is given by 
\begin{equation*}
    f(y_k \,|\, x_k, v_k, w_k) = \delta(y_k - x_k \, v_k - w_k) \enspace .
\end{equation*}
The conditional density $f(y_k, v_k, w_k \,|\, x_k)$ is given by
\begin{equation*}
    f(y_k, v_k, w_k \,|\, x_k) = \delta(y_k - x_k \, v_k - w_k) f_k^{vw}(v_k,w_k) \enspace .
\end{equation*}
Marginalization gives
\begin{equation*}
    \begin{aligned}
        f(y_k \,|\, x_k) 
        & = \int_\NewR \int_\NewR f(y_k, v_k, w_k \,|\, x_k) \, \mathrm{d} v_k \, \mathrm{d} w_k \\
        & = \int_\NewR \int_\NewR \delta(y_k - x_k \, v_k - w_k) 
        f_k^{vw}(v_k,w_k) \, \mathrm{d} v_k \, \mathrm{d} w_k \\
        & = \int_\NewR f_k^{vw}(v_k, y_k - x_k \, v_k) \, \mathrm{d} v_k \enspace .
    \end{aligned}
\end{equation*}
For $v_k$ and $w_k$ independent, i.e., $f_k^{vw}(v_k,w_k) = f_k^v(v_k) \, f_k^w(w_k)$, we obtain
\begin{equation*}
    f(y_k \,|\, x_k) = \int_\NewR f_k^v(v_k) \, f_k^w(y_k - x_k \, v_k) \, \mathrm{d} v_k \enspace .
\end{equation*}
For $f_k^v(v_k)$ and $f_k^w(w_k)$ Gaussian densities $f_k^v(v_k) \sim {\cal N}(v_k, m_k^v, \sigma_k^v)$ and $f_k^w(w_k) \sim {\cal N}(w_k, m_k^w, \sigma_k^w)$, the integral can be evaluated in closed-form as 
\begin{equation*}
    f(y_k \,|\, x_k) = \frac{1}{\sqrt{2 \pi} 
    \sqrt{\big( \sigma_k^v \big)^2 \, x_k^2 + \big( \sigma_k^w \big)^2}} 
    \exp\left\{ -\frac{1}{2} \frac{\big( y_k - m_k^v \, x_k - m_k^w \big)^2}{\big( \sigma_k^v \big)^2 \ x_k^2 
    + \big( \sigma_k^w \big)^2} \right\} \enspace .
\end{equation*}
For $m_k^v=0$ and $m_k^w=0$ this gives
\begin{equation*}
    f(y_k \,|\, x_k) = \frac{1}{\sqrt{2 \pi} \sqrt{\big( \sigma_k^v \big)^2 \, x_k^2 + \big( \sigma_k^w \big)^2}} 
    \exp\left\{ -\frac{1}{2} \frac{y_k^2}{\big( \sigma_k^v \big)^2 \, x_k^2 + \big( \sigma_k^w \big)^2} \right\} \enspace .
\end{equation*}
When the additive noise $w_k$ goes to zero, we obtain
\begin{equation*}
    f(y_k \,|\, x_k) = \frac{1}{\sqrt{2 \pi} \big( \sigma_k^v \big)^2 \, |x_k| } 
    \exp\left\{ -\frac{1}{2} \frac{y_k^2}{\big( \sigma_k^v \big)^2v \, x_k^2} \right\} \enspace ,
\end{equation*}
which is equivalent to \Eq{Eq_likelihoodMult} in \Ex{Ex_likelihoodMult} for Gaussian noise $f_k^v(.)$. When the multiplicative noise vanishes, which can be expressed as $f_k^v(v_k) = \delta(v_k-1)$ or $m_k^v=1$ and $\sigma_k^v \rightarrow 0$, we obtain
\begin{equation*}
    f(y_k \,|\, x_k) = \frac{1}{\sqrt{2 \pi} \sigma_k^w}
    \exp\left\{ -\frac{1}{2} \frac{(y_k - x_k)^2}{\big( \sigma_k^w \big)^2} \right\} \enspace ,
\end{equation*}
which is the result that is also directly obtained for the purely additive case.
\end{example}

%
%

\subsection{Time Update} \label{Subsec_ProbForm_TimeUpdate}

%
%
So far, we looked at a single measurement update, where a given measurement is used to update a prior estimate of the system state.
%
%
Now we consider the propagation of a given state estimate through a discrete-time nonlinear stochastic system
\begin{equation*}
    \vec{x}_{k+1} = \vec{a}_k(\vec{x}_k, \vec{w}_k)
\end{equation*}
with time step $k$, state $\vec{x}_k$, system noise $\vec{w}_k$, and nonlinear system function $\vec{a}_k(.,.)$. 
%
%
In analogy to the likelihood function as a probabilistic description of the measurement equation, we will use a probabilistic representation of the system equation. 
This so-called transition density is given by
\begin{equation*}
    f(\vec{x}_{k+1} \,|\, \vec{x}_k) \enspace .
\end{equation*}
It is calculated based on the system equation in a similar way as the likelihood function is determined based on the measurement equation.

%
%

\section{Zakai-type Distributed Ordinary Differential Equations in Artificial Time} 
\label{Sec_Derivation_Zakai_DODE}

%
%
In this section, we will derive distributed ordinary differential equations (DODEs) that smoothly morph one density into another. 
Here, we will focus on infinite-dimensional representations, i.e., function spaces, with corresponding solutions derived in \Sec{Sec_Exact_Soln} for some special cases.
Finite-dimensional spaces will be considered in \Sec{Sec_Parametric_Soln} and \Sec{Sec_Nonparametric_Soln}. 
Parametric solutions are derived in \Sec{Sec_Parametric_Soln}, while \Sec{Sec_Nonparametric_Soln} covers non-parametric solutions.

%
%
In the following, we omit the time index $k$ for simplicity.

%
%

\subsection{Key Idea}

%
%
The most general setup is as follows. 
We are given two densities $f_0(\vec{x})$ and $f_1(\vec{x})$. 
The first density $f_0(\vec{x})$ is not necessarily known, but an approximation $f_0(\vec{x},\vec{\eta}_0)$ is given, where $\vec{\eta}_0$ is a parameter vector. 
The second density, the target, may only be implicitly known or only be given at certain points. 
Our goal is to find an approximation $f_1(\vec{x},\vec{\eta}_1)$.
%
%
Finding the parameter vector $\vec{\eta}_1$ typically is a complicated optimization problem.

%
%
The key idea for efficiently obtaining $\vec{\eta}_1$ is to first find a smooth progression for morphing $f_0(\vec{x})$ into $f_1(\vec{x})$ depending on an artificial time parameter $\gamma \in [0,1]$, i.e.,
\begin{equation*}
    f_\gamma(\vec{x}) = 
    \begin{cases}
        f_0(\vec{x}) & \gamma=0 \\
        \text{nice interpolation} & \gamma \in (0,1) \\
        f_1(\vec{x}) & \gamma=1
    \end{cases}
    \enspace .
\end{equation*}
%
%
Second, this progression is expressed as a DODE
\begin{equation*}
    \dot{f}_\gamma(\vec{x}) = \frac{\partial \, f_\gamma(\vec{x})}{\partial \, \gamma}
    = a_\gamma(\vec{x}) \, f_\gamma(\vec{x}) + b_\gamma(\vec{x})
\end{equation*} 
depending on the artificial time $\gamma \in [0,1]$ and the state $\vec{x}$. $a_\gamma(\vec{x})$ and $b_\gamma(\vec{x})$ are suitable functions, that will be derived in the following for specific problems.
%
%
The initial condition is $f_{\gamma=0}(\vec{x})=f_0(\vec{x})$ and for $\gamma=1$, we obtain the desired density $f_1(\vec{x})$.
%
%
Third, this DODE is transferred to the approximate parametrized densities $f_0(\vec{x},\vec{\eta}_0)$ and $f_1(\vec{x},\vec{\eta}_1)$.
%
%
\begin{remark}
It is important to note the following two properties of the DODE:
\begin{itemize}
\item The derived evolution equation is \emph{not} a partial differential equation, but rather an ordinary differential equation for all values of the state $\vec{x}$. Hence, it is called a distributed ordinary differential equation (DODE).
\item We consider the evolution of \emph{unnormalized} densities. This is the reason, why we call them Zakai-type DODEs.
\end{itemize}
These two properties make the solution for $f_\gamma(\vec{x})$, $\gamma \in [0,1]$ much simpler compared to state of the art approaches.
\end{remark}
%
%
Fourth, by averaging the DODE resulting from inserting the approximate densities over the domain $\vec{x}$, we finally obtain a system of ordinary differential equations for the parameter vector
\begin{equation*}
    \dot{\vec{\eta}}_\gamma = \mat{M}(\vec{\eta}_\gamma) \, \vec{\eta}_\gamma
\end{equation*} 
with
\begin{equation*}
    \vec{\eta}_\gamma = 
    \begin{cases}
        \vec{\eta}_0 & \gamma=0 \\
        \vec{\eta}_1 & \gamma=1
    \end{cases}
    \enspace .
\end{equation*}
$\mat{M}(\vec{\eta}_\gamma)$ is a matrix depending on the functions $a_\gamma(\vec{x})$ and $b_\gamma(\vec{x})$ and the selected approximate representation $f_\gamma(\vec{x},\vec{\eta}_\gamma)$.

%
%

\subsection{DODE for Morphing from One Density to Another Density}

%
%
The simplest progression to move from $f_0$ to $f_1$ is given by
\begin{equation*}
    f_\gamma(\vec{x}) = \big( 1-g_\gamma \big) f_0(\vec{x}) + g_\gamma \, f_1(\vec{x}) 
    = f_0(\vec{x}) + g_\gamma \big( f_1(\vec{x}) - f_0(\vec{x}) \big) \enspace ,
\end{equation*}
with  $g_\gamma \in [0,1]$, i.e., $g_0=0$ and $g_1=1$, $g_\gamma$ continuous, and $g_\gamma$ strictly increasing.
%
%
Taking the derivative with respect to $\gamma$ gives
\begin{equation*}
    \dot{f}_\gamma(\vec{x}) = \dot{g}_\gamma \big( f_1(\vec{x}) - f_0(\vec{x}) \big)
\end{equation*}
or with
\begin{equation*}
    f_1(\vec{x}) - f_0(\vec{x}) = \frac{1}{g_\gamma} \big( f_\gamma(\vec{x}) - f_0(\vec{x}) \big)
\end{equation*}
the DODE
\begin{equation*}
    \dot{f}_\gamma(\vec{x}) = \frac{\dot{g}_\gamma}{g_\gamma} \big( f_\gamma(\vec{x}) - f_0(\vec{x}) \big)
    \enspace ,
\end{equation*}
where we have to take care of the singularity due to $g_\gamma$ when $\gamma=0$.
%
%
\begin{remark}
The function $g_\gamma$ determines the rate of change in the densities when $\gamma$ varies from $0$ to $1$. Of course, we could use $g_\gamma=\gamma$ when the expected change is relatively homogeneous over $\gamma$ anyway. However, when foreseeable variations in the rate of change occur, an appropriate selection of $g_\gamma$ can simplify the life of the DODE solver.
\end{remark}

%
%

\subsection{DODE for Measurement Update}

In the case of a measurement update according to \Eq{Eq_Posterior_normalized}, we set $f_0(\vec{x}) = f_p(\vec{x})$ and $f_1(\vec{x})= f_e(\vec{x})$. Omitting the normalization constant in \Eq{Eq_Posterior_normalized}, the relation between the two densities is given by
\begin{equation*}
    f_e(\vec{x}) = f_L(\vec{x}) \, f_p(\vec{x}) \enspace .
\end{equation*}
We now introduce a progressive likelihood function $f_L(\vec{x},\gamma)$ with the property
\begin{equation*}
    f_L(\vec{x},\gamma) =
    \begin{cases}
        1 & \gamma=0 \\
        f_L(\vec{x}) & \gamma=1
    \end{cases}   
\end{equation*}
and we obtain
\begin{equation*}
    f_e(\vec{x},\gamma) = f_L(\vec{x},\gamma) \, f_p(\vec{x}) \enspace .
\end{equation*}
%
%
Taking the derivative with respect to $\gamma$ gives
\begin{equation*}
    \dot{f}_e(\vec{x},\gamma) = \dot{f}_L(\vec{x},\gamma) \, f_p(\vec{x})
\end{equation*}
or
\begin{equation*}
    \dot{f}_e(\vec{x},\gamma) = \frac{\dot{f}_L(\vec{x},\gamma)}{f_L(\vec{x},\gamma) } 
    \, f_e(\vec{x}, \gamma) \enspace ,
\end{equation*}
which is the desired DODE.

%
%
A straightforward progression is
\begin{equation*}
    f_L(\vec{x},\gamma) = \left[ f_L(\vec{x}) \right]^{g(\gamma)} \enspace .
\end{equation*}
Its derivative is
\begin{equation*}
    \begin{aligned}
    \dot{f}_L(\vec{x},\gamma) 
    & = \frac{\partial}{\partial \, \gamma} \left\{ \left[ 
        \exp\Big( \log\big( f_L(\vec{x}) \big) \Big) \right]^{g(\gamma)} \right\} \\[2mm]
    & = \frac{\partial}{\partial \, \gamma} \left\{ 
    \exp\Big( g(\gamma) \,\log\big( f_L(\vec{x}) \big) \Big) \right\} \\[2mm]
    & = \log\big( f_L(\vec{x}) \big) \, \dot{g}(\gamma) \, f_L(\vec{x})
    \enspace .
    \end{aligned}
\end{equation*}
%
%
The DODE for this progression is then
\begin{equation*}
    \dot{f}_e(\vec{x},\gamma) = \log\big( f_L(\vec{x}, \gamma) \big) \, \dot{g}(\gamma)
    \, f_e(\vec{x}, \gamma) \enspace .
\end{equation*}
%

%
%
For exponential families, we obtain simpler expressions due to the logarithm taken above. For the special case of additive Gaussian measurement noise according to \SubSec{SubSubSec_ProbForm_AdditiveNoise} and $g(\gamma)=\gamma$, the likelihood function is
\begin{equation*}
    f_L(\vec{x}, \gamma) 
    = \exp\left( - \, \gamma \, \frac{1}{2}
    \big( \vec{y}-\vec{h}(\vec{x}) \big)^\top \mat{C}_v^{-1} \big( \vec{y}-\vec{h}(\vec{x}) \big) 
    \right)
    \enspace ,
\end{equation*}
and in the scalar case
\begin{equation*}
    f_L(x, \gamma) 
    = \exp\left( - \, \gamma \, \frac{1}{2} \frac{(y-h(x))^2}{\sigma_v^2} \right)
    \enspace .
\end{equation*}
The corresponding DODE is
\begin{equation*}
    \dot{f}_e(\vec{x}, \gamma)
    = - \frac{1}{2} 
    \big( \vec{y}-\vec{h}(\vec{x}) \big)^\top \mat{C}_v^{-1} \big( \vec{y}-\vec{h}(\vec{x}) \big) 
    \, f_e(\vec{x}, \gamma) 
\end{equation*}
and in the scalar case
\begin{equation} \label{Eq_DODE}
    \dot{f}_e(x, \gamma)
    = - \frac{1}{2} \frac{[y-h(x)]^2}{\sigma_v^2}
    \, f_e(x, \gamma) \enspace .
\end{equation}
%

%
%

\section{Zakai Equation: Exact Solution} \label{Sec_Exact_Soln}

The distributed ordinary differential equation (DODE) in \Eq{Eq_DODE} can be solved exactly in a few simple (but interesting) cases. 
In general, however, approximate solution methods have to be employed. 
For the special case of a linear measurement equation, the solution procedure is shown in the next example. 
Of course, we obtain the expected result for the posterior density that in this case could have been easily obtained by directly solving \Eq{Eq_Posterior_normalized}.

\begin{example}
    We consider a prior zero-mean Gaussian density
    \begin{equation*} 
        f_e(x, \gamma=0)=f_p(x)={\cal N}(x,0,\sigma_p^2)
        = \frac{1}{\sqrt{2 \pi} \sigma_p^2}
        \exp\left( - \frac{1}{2} \frac{x^2}{\sigma_p^2}\right)
    \end{equation*}
    with standard deviation $\sigma_p$. 
    The measurement equation is assumed to be simply
    \begin{equation*}
        y=x+v \enspace ,
    \end{equation*} 
    with measurement $y$ and measurement noise $v ~ \sim {\cal N}(v, 0, \sigma_v)$, which corresponds to the likelihood function
    \begin{equation*}
        f_L(x)=\exp\left( - \frac{1}{2} 
        \frac{(y-x)^2}{\sigma_v^2} \right) \enspace ,
    \end{equation*} 
    where we set $y=0$ for simplicity. 
    According to \Eq{Eq_DODE}, the DODE is given by
    \begin{equation} \label{Eq_ExactSolnDODE}
        \dot{f}_e(x, \gamma)
        = - \frac{1}{2} \frac{x^2}{\sigma_v^2} \, f_e(x,\gamma) \enspace .
    \end{equation}
    For the solution of the DODE, we use the ansatz
    \begin{equation*}
        \bar{f}_e(x, \gamma) = k_1 \, \exp(k_2 \, \gamma \, x^2)
    \end{equation*}
    with derivative with respect to $\gamma$ given by
    \begin{equation*}
        \dot{\bar{f}}_e(x, \gamma) = k_1 \, k_2 \, x^2 \,
        \exp(k_2 \, \gamma \, x^2)
        = k_2 \, x^2 \, f_e(x, \gamma) \enspace .
    \end{equation*}
    Inserting $\bar{f}_e(x, \gamma)$ and $\dot{\bar{f}}_e(x, \gamma)$ into the DODE gives
    \begin{equation*}
    k_2 \, x^2 \, f_e(x, \gamma) 
    = - \frac{1}{2} \frac{x^2}{\sigma_v^2} f_e(x, \gamma)
    \end{equation*}
    and it follows that
    \begin{equation*}
        k_2 = - \frac{1}{2} \frac{1}{\sigma_v^2} \enspace .
    \end{equation*}
    Using the initial condition
    \begin{equation*}
        \dot{\bar{f}}_e(x, \gamma=0) = k_1
        = f_e(x, \gamma=0)
        = \frac{1}{\sqrt{2 \pi} \sigma_p^2}
        \exp\left( - \frac{1}{2} \frac{x^2}{\sigma_p^2}\right)
    \end{equation*}
    gives
    \begin{equation*}
        \begin{aligned}
            \bar{f}_e(x, \gamma) & = 
            \frac{1}{\sqrt{2 \pi} \sigma_p^2}
            \exp\left( - \frac{1}{2} \frac{x^2}{\sigma_p^2}\right)
            \exp\left( - \gamma \frac{1}{2} \frac{x^2}{\sigma_v^2} \right) \\
            & = \frac{1}{\sqrt{2 \pi} \sigma_p^2}
            \exp\left( - \frac{1}{2} 
            \left( \frac{x^2}{\sigma_p^2} 
            + \gamma \frac{x^2}{\sigma_v^2} \right)
            \right) \enspace .
        \end{aligned}
    \end{equation*}
    This can be written as
    \begin{equation} \label{Eq_ExactSolnGauss}
        \bar{f}_e(x, \gamma) 
        = \frac{1}{\sqrt{2 \pi} \sigma_p^2}
        \exp\left( - \frac{1}{2} \frac{x^2}{\sigma_e^2} \right)
    \end{equation}
    with
    \begin{equation*}
        \sigma_e^2 = \frac{1}{\frac{1}{\sigma_p^2}
        + \gamma \frac{1}{\sigma_v^2}} \enspace .
    \end{equation*}
    Obviously, we get the same result as solving \Eq{Eq_Posterior_normalized} directly for this special case of a linear system.
    Please note that the posterior density in \Eq{Eq_ExactSolnGauss} \emph{is unnormalized} (in this case, normalized with respect to the prior density) as the DODE in \Eq{Eq_ExactSolnDODE} describes the evolution of the unnormalized density starting at the (in this case) normalized prior.
\end{example} 

%
%

\section{Parametric Solution of Zakai Equation} \label{Sec_Parametric_Soln}

%
%

For simplifying the exposition, we begin with the scalar DODE \Eq{Eq_DODE} repeated here for convenience
\def\rhsInner{\frac{[y-h(x)]^2}{\sigma_v^2}}
\def\rhs{ - \frac{1}{2} \, \rhsInner}
\begin{equation*}
    \dot{f}_e(x, \gamma) = \rhs f_e(x, \gamma) \enspace .
\end{equation*}

%
%
\def\etaVec{\vec{\eta}_e(\gamma)}
For solving this DODE, we plug in a parametric representation $f_e(x, \gamma) = f_e(x, \vec{\eta}_e(\gamma))$ with parameter vector $\etaVec$, which gives
\def\feArgsEta{\big( x, \etaVec \big)}
\def\feEta{f_e \feArgsEta}
\def\dotFeEta{\dot{f}_e \feArgsEta}
\begin{equation} \label{Eq_DODE_scalar_eta}
    \dotFeEta = \rhs \feEta \enspace .
\end{equation}
%
%
In order to calculate the desired parameter vector $\etaVec$ for $\gamma=1$, the DODE has to be solved for $\gamma \in [0,1]$ and for all states $x$ simultaneously. 
%
%
However, for a given parametric representation, the DODE typically cannot be exactly solved over all states $x$. 
%
%
Hence, we have to find a parameter vector $\etaVec$ in such a way that some average distance between the left-hand side and the right-hand side of \Eq{Eq_DODE_scalar_eta} is minimized for all $\gamma$.
%
%
We will pursue two options for minimizing an average distance: (1) a continuous version based on the squared integral distance and (2) a discrete version based on distances at collocation points.
%
%
Both options will be used for different parametric representations.

%
%

\subsection{Parametric Solution of Zakai Equation: Gaussian Posteriors} \label{Sec_Parametric_Soln_Gauss}

%
%

As a specific parametric representation of the posterior density, we assume that the state pdf can be represented by an (unnormalized) posterior Gaussian density, i.e., 
\def\feRhs{k_e(\gamma) \exp\left( -\frac{1}{2}
    \frac{[x-m_e(\gamma)]^2}{\sigma_e^2(\gamma)} \right)}
\begin{equation} \label{Eq_GaussPost}
    \feEta = \feRhs \enspace ,
\end{equation}
with
\begin{equation*}
    \etaVec = \begin{bmatrix} k_e(\gamma) & m_e(\gamma) & \sigma_e(\gamma) \end{bmatrix}^\top
\end{equation*}
depending on the artificial time $\gamma$.

%
%

\begin{remark}
The factor $k_e(\gamma)$ is not used for normalizing $\feEta$. 
For the start of the progression, i.e., $\gamma=0$, it can be initialized with any convenient value such as $1$ or the normalization constant of the prior density.
During the progression, it is then used to take care of the change in probability mass due to the multiplication with the likelihood function. 
When the progression reaches $\gamma=1$, $k_e(\gamma)$  is generally discarded and the density $\feEta$ properly normalized.
\end{remark}

%
%

The derivative of $\feEta$ with respect to $\gamma$ is given by
\def\lhsInner{\frac{1}{k_e(\gamma)} \dot{k}_e(\gamma)%
    + \frac{x-m_e(\gamma)}{\sigma_e^2(\gamma)} \dot{m}_e(\gamma)%
    + \frac{[x-m_e(\gamma)]^2}{\sigma_e^3(\gamma)} \dot{\sigma}_e(\gamma)}
\def\lhs{\left( \lhsInner \right) \feEta }
\begin{equation} \label{Eq_GaussPostDiff}
    \dotFeEta = \lhs \enspace .
\end{equation}
%
%
Plugging \Eq{Eq_GaussPostDiff} into \Eq{Eq_DODE_scalar_eta} results in a parametric DODE
\begin{equation} \label{Eq_DODE_scalar_Gauss}
    \lhs = \rhs \feEta \enspace .
\end{equation}
%
%
Our goal is to solve the parametric DODE \Eq{Eq_DODE_scalar_Gauss} for $\gamma \in [0,1]$ in order to obtain the values of $k_e(\gamma)$, $m_e(\gamma)$, and $\sigma_e(\gamma)$ for $\gamma = 1$.

%
%

\subsubsection{Squared Integral Solution} \label{SubSec_GaussSquaredIntegral}

For solving the parametric DODE \Eq{Eq_DODE_scalar_Gauss}, we first employ a squared integral distance measure
\def\fe2gamma{f_e^2\big( x, \etaVec \big)}
\begin{equation} \label{Eq_SqIntDist}
    D = \int_\NewR \Bigg(
        \underbrace{\lhsInner + \frac{1}{2} \, \rhsInner}_{\Delta(x, \gamma)} 
    \Bigg)^2 \fe2gamma \, \mathrm{d}x \enspace .
\end{equation}
Taking the derivative of $D$ with respect to $\dot{k}_e(\gamma)$, $\dot{m}_e(\gamma)$, and $\dot{\sigma}_e(\gamma)$ gives
\begin{equation*}
    \begin{aligned}
        D_k = \frac{\partial D}{\partial \dot{k}_e(\gamma)} 
        & = 2 \, \int_\NewR \Delta(x, \gamma) \, \frac{1}{k_e(\gamma)} 
        \, \fe2gamma \,\mathrm{d}x \enspace , \\
        D_m = \frac{\partial D}{\partial \dot{m}_e(\gamma)} 
        & = 2 \, \int_\NewR \Delta(x, \gamma) \, \frac{x-m_e(\gamma)}{\sigma_e^2(\gamma)} 
        \, \fe2gamma \,\mathrm{d}x \enspace , \\
        D_\sigma = \frac{\partial D}{\partial \dot{\sigma}_e(\gamma)} 
        & = 2 \, \int_\NewR \Delta(x, \gamma) \, \frac{(x-m_e(\gamma))^2}{\sigma_e^3(\gamma)} 
        \, \fe2gamma \,\mathrm{d}x \enspace .
    \end{aligned}
\end{equation*}
The parameters $k_e(\gamma)$, $m_e(\gamma)$, and $\sigma_e(\gamma)$ corresponding to the minimum of $D$ in \Eq{Eq_SqIntDist} for all $\gamma$ is obtained by setting $D_k$, $D_m$, and $D_\sigma$ to zero, which gives
\def\qEta{\vec{q}\big( \etaVec \big)}
\begin{equation} \label{Eq_ODE_P_eta_q}
    \mat{P}\big( \etaVec \big) \, \cdot \, \dot{\vec{\eta}}_e(\gamma) = \qEta
\end{equation}
with 
\begin{equation*}
    \dot{\vec{\eta}}_e(\gamma) = \begin{bmatrix} \dot{k}_e(\gamma), \dot{m}_e(\gamma), \dot{\sigma}_e(\gamma) \end{bmatrix}^\top
\end{equation*}
and
\def\I#1{I\left( #1 \right)}
\begin{equation} \label{Eq_P_Gauss}
    \mat{P}\left( \etaVec \right) =
    \begin{bmatrix}
        \I{ \frac{1}{k_e^2(\gamma)} } &
        \I{ \frac{1}{k_e(\gamma)} \frac{x-m_e(\gamma)}{\sigma_e^2(\gamma)} } &
        \I{ \frac{1}{k_e(\gamma)} \frac{(x-m_e(\gamma))^2}{\sigma_e^3(\gamma)} } \\[8mm]
        \I{ \frac{1}{k_e(\gamma)} \frac{x-m(\gamma)}{\sigma_e^2(\gamma)} } &         
        \I{ \frac{\left( x-m_e(\gamma) \right)^2}{\sigma_e^4(\gamma)} } &
        \I{ \frac{\left( x-m_e(\gamma) \right)^3}{\sigma_e^5(\gamma)} } \\[8mm]
        \I{ \frac{1}{k_e(\gamma)} \frac{(x-m_e(\gamma))^2}{\sigma_e^3(\gamma)} } &
        \I{ \frac{\left( x-m_e(\gamma) \right)^3}{\sigma_e^5(\gamma)} } &
        \I{ \frac{\left( x-m_e(\gamma) \right)^4}{\sigma_e^6(\gamma)} }
    \end{bmatrix} \enspace ,
\end{equation}
where $\I{.}$ is defined as
\begin{equation*}
    \I{ \text{g}(x) } = \int_\NewR \text{g}(x)\, \fe2gamma \, \mathrm{d}x \enspace .
\end{equation*}
The right-hand side of \Eq{Eq_ODE_P_eta_q} is given by
\begin{equation} \label{Eq_q_integral}
    \qEta = - \frac{1}{2}
    \begin{bmatrix}
        \displaystyle
        \int_\NewR \rhsInner \, \frac{1}{k_e(\gamma)}                      \, \fe2gamma \, \mathrm{d}x \\[8mm]
        \displaystyle
        \int_\NewR \rhsInner \, \frac{x-m_e(\gamma)}{\sigma_e^2(\gamma)}     \, \fe2gamma \, \mathrm{d}x \\[8mm]
        \displaystyle
        \int_\NewR \rhsInner \, \frac{(x-m_e(\gamma))^2}{\sigma_e^3(\gamma)} \, \fe2gamma \, \mathrm{d}x
    \end{bmatrix}
    \enspace .
\end{equation}

For solving the integral expressions in \Eq{Eq_P_Gauss}, we use the following moment expressions
\begin{equation} \label{Eq_Integral_I}
    \I{ \big( x-m_e(\gamma) \big)^n } = \int_\NewR \big( x-m_e(\gamma) \big)^n \, \fe2gamma \, \mathrm{d}x
    = 
    \begin{cases}
        \sqrt{\pi} \, k_e^2(\gamma) \, \frac{(n-1)(n-3) \cdots 1}{2^{n/2}} 
            \sigma^{n+1}(\gamma) & \text{ for $n$ even} \\
        0 & \text{ for $n$ odd}
    \end{cases} \enspace .
\end{equation}
For the cases $n=0, \ldots, 4$, expressions are given in \Fig{Fig_Integral_I}.
\begin{figure*}
\begin{alignat*}{2} 
    n=0: \;\; & \I{ \big( x-m_e(\gamma) \big)^0 } = \int_\NewR f_e^2(\vec{x}, \gamma) & \, 
        = \, & \sqrt{\pi} \, k_e^2(\gamma) \, \sigma_e(\gamma)               \enspace , \\
    n=1: \;\; & \I{ \big( x-m_e(\gamma) \big)^1 } = \int_\NewR (x-m_e(\gamma)) \, \fe2gamma & \, 
        = \, & 0 \enspace , \\
    n=2: \;\; & \I{ \big( x-m_e(\gamma) \big)^2 } = \int_\NewR (x-m_e(\gamma))^2 \, \fe2gamma  & \, 
        = \, & \frac{1}{2} \sqrt{\pi} \, k_e^2(\gamma) \, \sigma^3(\gamma) \enspace , \\ 
    n=3: \;\; & \I{ \big( x-m_e(\gamma) \big)^3 } = \int_\NewR (x-m_e(\gamma))^3 \, \fe2gamma  & \, 
        = \, & 0 \enspace , \\
    n=4: \;\; & \I{ \big( x-m_e(\gamma) \big)^4 } = \int_\NewR (x-m_e(\gamma))^4 \, \fe2gamma  & \, 
        = \, & \frac{3}{4} \sqrt{\pi} \, k_e^2(\gamma) \, \sigma_e^5(\gamma) \enspace .
\end{alignat*}
\caption{Special cases $n=0, \ldots, 4$ for the integral $\I{ \big( x-m_e(\gamma) \big)^n }$ in \Eq{Eq_Integral_I}.}
\label{Fig_Integral_I}
\end{figure*}
As a result, we obtain
\begin{equation*}
    \mat{P}\left( \etaVec \right) = \sqrt{\pi}
    \begin{bmatrix}
        \sigma_e(\gamma)           & 0              & \frac{1}{2} \, k_e(\gamma)                 \\[3mm]
        0                          & \frac{1}{2} \, \frac{k_e^2(\gamma)}{\sigma_e(\gamma)}  & 0    \\[3mm]
        \frac{1}{2} \, k_e(\gamma) & 0   & \frac{3}{4} \, \frac{k_e^2(\gamma)}{\sigma_e(\gamma)}
    \end{bmatrix} \enspace .
\end{equation*}
%

%
%

\paragraph{Analytic Solution for $\qEta$ in \Eq{Eq_q_integral}}

%
%
In many interesting cases, an analytic solution for $\qEta$ can be found. 
%
%
The following example gives the solution for the case of a linear system. This serves as a sanity check, as in this case, the exact solution is known. 
\begin{example}
    We consider a scalar linear measurement equation, i.e., $h(x)$ in \Eq{Eq_Measurement_additive} is now given as $h(x) = H \, x$. In this case, we obtain
    \begin{equation*}
        \qEta = \frac{\sqrt{\pi}}{2 \, \sigma_v^2}
        \begin{bmatrix}
            \displaystyle
            - \frac{1}{2} k_e(\gamma) \, \sigma_e(\gamma) \, \left\{ H^2 \sigma_e^2(\gamma) + 2 \big( y - H \, m_e(\gamma) \big)^2 \right\} \\[3mm]
            \displaystyle
            H \, k_e^2(\gamma) \, \sigma_e(\gamma) \, \big( y - H \, m_e(\gamma) \big) \\[3mm]
            \displaystyle
            - \frac{1}{4} k_e^2(\gamma) \, \left\{ 3 H^2 \sigma_e^2(\gamma) + 2 \big( y - H \, m_e(\gamma) \big)^2 \right\}
        \end{bmatrix}
    \end{equation*}
    as the right-hand side of \Eq{Eq_ODE_P_eta_q}.
\end{example}

%
%

For solving the ODE \Eq{Eq_ODE_P_eta_q}, we use a standard ODE solver. We avoid the explicit inversion of the matrix $\mat{P}\big( \vec{\eta}(\gamma) \big)$ by interpreting \Eq{Eq_ODE_P_eta_q} as a linear equation that is solved for $\dot{\vec{}\eta}(\gamma)$.
%
%
The ODE solver is accelerated by using the Jacobian matrix of $\qEta$ given by
\begin{equation*}
    J\big( \etaVec \big)
    = \frac{\partial \vec{q}^\top\big( \etaVec \big)}
           {\partial \etaVec } \enspace .
\end{equation*}

%
%

There are other interesting nonlinear functions $h(x)$, where the analytic solution for $\qEta$ in \Eq{Eq_q_integral} is possible. This includes polynomials, trigonometric functions, and exponential functions.

%
%

\paragraph{Numeric Solution for $\qEta$ in \Eq{Eq_q_integral}}

When an analytic solution is not possible, \Eq{Eq_q_integral} has to be evaluated numerically. 
An elegant quadrature rule is obtained by replacing $\fe2gamma$ by its Dirac mixture approximation given by
\begin{equation*}
    f_e^2(x, \gamma) 
    \approx \sum_{i=0}^N w_i(\gamma) \, \delta\big( x-x_i(\gamma) \big)
    \enspace .
\end{equation*}
Inserting this approximation into \Eq{Eq_q_integral} gives
\begin{equation*}
    \vec{q}\left( \vec{\eta}(\gamma) \right) \approx - \frac{1}{2}
    \sum_{i=0}^N w_i(\gamma) \, 
    \frac{\big[ y-h\big( x_i(\gamma) \big) \big]^2}{\sigma_v^2}
    \begin{bmatrix}
        \frac{1}{k_e(\gamma)} \\[8mm]
        \frac{x_i(\gamma)-m_e(\gamma)}{\sigma_e^2(\gamma)} \\[8mm]
        \frac{(x_i(\gamma)-m_e(\gamma))^2}{\sigma_e^3(\gamma)}
    \end{bmatrix}
    \enspace .
\end{equation*}
%

%
%

\begin{remark}
Of course, we could now ask ourselves why this numerical procedure is simpler than just performing a numeric integration of the basic update equation \Eq{Eq_Posterior_normalized} for calculating the desired posterior moments. 
First, the square of the progressive posterior $f_e^2(x, \etaVec)$ is always narrower than the prior $f_p(x)$ even for $\gamma=0$ and gets narrower with increasing $\gamma$. Hence, we have to integrate over a smaller region. 
Second, we do not have to integrate over the likelihood as in \Eq{Eq_Posterior_normalized}, we only have to calculate (nonlinear) moments of $f_e^2(x, \etaVec)$. 
Third, this procedure also works for Gaussian Mixture (GM) posteriors, not only for Gaussian posteriors. For GM posteriors, moments of \Eq{Eq_Posterior_normalized} would not be helpful for obtaining the posterior mixture parameters.
\end{remark}

%
%

\subsubsection{Solution at Collocation Points} \label{SubSec_GaussCollocation}

%
%
In \SubSec{SubSec_GaussSquaredIntegral}, the distributed ordinary differential equation \Eq{Eq_DODE_scalar_Gauss} was solved ``on average'' over the state domain by minimizing a squared integral distance between its left-hand side and its right-hand side for calculating $\dot{\vec{\eta}}_e(\gamma)$.
%
%
Here, we will solve the DODE \Eq{Eq_DODE_scalar_Gauss} at selected discrete points only. There are several ways of selecting these points. The natural solution is to use a Dirac mixture approximation of $\feEta$
\begin{equation*}
    \feEta
    \approx \sum_{i=0}^C w_i(\gamma) \, \delta\big( x-x_i(\gamma) \big)
    \enspace ,
\end{equation*}
with $x_i(\gamma)$ the collocation points for $i=1,\ldots,C$. The weights $w_i(\gamma)$ are typically selected to be equal. This automatically places more points where the density is large, putting a higher priority to large density regions.

%
%

By plugging these collocation points into \Eq{Eq_DODE_scalar_Gauss}, we obtain
\def\lhs{\frac{1}{k_e(\gamma)} \dot{k}_e(\gamma)%
    + \frac{x_i(\gamma)-m_e(\gamma)}{\sigma_e^2(\gamma)} \dot{m}_e(\gamma)%
    + \frac{[x_i(\gamma)-m_e(\gamma)]^2}{\sigma_e^3(\gamma)} \dot{\sigma}_e(\gamma)}
\def\rhs{- \frac{1}{2} \, \frac{[y-h(x_i(\gamma))]^2}{\sigma_v^2}}
\begin{equation*}
    \lhs = \rhs
\end{equation*}
for $i=1,\ldots,C$. In vector-matrix-form, in analogy to \Eq{Eq_ODE_P_eta_q}, this gives
\begin{equation*}
    \mat{P}\big( \etaVec \big) \, \cdot \, \dot{\vec{\eta}}_e(\gamma) = \qEta
\end{equation*}
with
\begin{equation*}
    \mat{P}\big( \etaVec \big) =
    \begin{bmatrix}
        \frac{1}{k_e(\gamma)} 
        & \frac{x_1(\gamma)-m_e(\gamma)}{\sigma_e^2(\gamma)} 
        & \frac{[x_1(\gamma)-m_e(\gamma)]^2}{\sigma_e^3(\gamma)} \\[1mm]
        \frac{1}{k_e(\gamma)} 
        & \frac{x_2(\gamma)-m_e(\gamma)}{\sigma_e^2(\gamma)} 
        & \frac{[x_2(\gamma)-m_e(\gamma)]^2}{\sigma_e^3(\gamma)} \\[1mm]
        \vdots & \vdots & \vdots \\[1mm]
        \frac{1}{k_e(\gamma)} 
        & \frac{x_i(\gamma)-m_e(\gamma)}{\sigma_e^2(\gamma)} 
        & \frac{[x_i(\gamma)-m_e(\gamma)]^2}{\sigma_e^3(\gamma)} \\[1mm]
        \vdots & \vdots & \vdots \\[1mm]
        \frac{1}{k_e(\gamma)} 
        & \frac{x_C(\gamma)-m_e(\gamma)}{\sigma_e^2(\gamma)} 
        & \frac{[x_C(\gamma)-m_e(\gamma)]^2}{\sigma_e^3(\gamma)}
    \end{bmatrix}
\end{equation*}
and
\begin{equation*}
    \qEta =
    - \frac{1}{2} \, \frac{1}{\sigma_v^2}
    \begin{bmatrix}
        [y-h(x_1(\gamma))]^2 \\[1mm]
        [y-h(x_2(\gamma))]^2 \\[1mm]
        \vdots \\[1mm]
        [y-h(x_i(\gamma))]^2 \\[1mm]
        \vdots \\[1mm]
        [y-h(x_C(\gamma))]^2
    \end{bmatrix} \enspace .
\end{equation*}
%

%
%

\section{Nonparametric Solution of Zakai Equation} \label{Sec_Nonparametric_Soln}

%
%
Here, we extend the previous ideas to the case of nonparametric solutions. 
By ``nonparametric'' we mean that we directly use a set of samples that is updated during the measurement step. 
No fitting of these samples to a parametric density is performed.

%
%
We are given a set of $L_p$ prior sample vectors $\vec{x}_{p,i}$, $i=1, \ldots, L_p$ arranged in a matrix according to
\begin{equation*} 
    \mat{X}_p = \begin{bmatrix} \vec{x}_{p,1}, \vec{x}_{p,2}, \ldots, \vec{x}_{p,i}, \ldots, \vec{x}_{p,L_p} \end{bmatrix} \enspace.
\end{equation*}
%
%
Our goal is to find a posterior set of $L_e$ sample vectors $\vec{x}_{e,i}(\gamma)$, $i=1, \ldots, L_e$ with
\begin{equation*} 
    \mat{X}_e(\gamma) = \begin{bmatrix} \vec{x}_1(\gamma), \vec{x}_2(\gamma), 
    \ldots, \vec{x}_i(\gamma), \ldots, \vec{x}_{L_e}(\gamma) \end{bmatrix} \enspace.
\end{equation*}
%
%
For simplicity, we start with the 
%
%
case of equally weighted samples in the prior and the posterior set $\mat{X}_p$ and $\mat{X}_e(\gamma)$. 
%
%
We also assume that the number of components does not change during the update, so $L=L_p=L_e$. 
%
%
Furthermore, we assume a scalar state domain.

%
%
The key idea is to reconstruct the values of the density underlying the given samples at the sample locations. 
A simple density-spacings estimator is used for that purpose.
%
%
An estimate of the density underlying the sample set $\mat{X}_e(\gamma)$ is given by
\begin{equation*} 
    f_e(x_i, \gamma) = \frac{1}{2} \frac{w_i}{ | x_i(\gamma) - x_j(\gamma) | } \enspace ,
\end{equation*}
where $w_i=1/L$ and $j$ is the index of the nearest neighbor sample to sample $i$. We write $j=\text{NN}(i)$. 
%
%
\begin{remark}
%
%
Considering the density only at the given $L$ sample locations would directly give us $L$ ODEs for the desired posterior locations.
%
%
However, even in the one-dimensional case we need $L+1$ equations as we have to propagate the $L$ locations plus some factor that takes care of the fact that we consider unnormalized densities.

%
%
Of course, in higher-dimensional state spaces, say in $D$ dimensions, $L \cdot D$ parameters are required for representing $L$ samples plus one parameter for the factor.
%
%
As in the scalar case, considering density values at the component locations still only gives us $L$ equations for, now, $L \cdot D + 1$ parameters.

%
%
In order to obtain more equations than unknowns, we perform first a continuous interpolation between the density values leading to a DODE that is continuous over the state space $\cal X$.
%
%
Second, this continuous DODE is evaluated at $N$ carefully selected collocation points with $N>L \cdot D + 1$.
%
%
resulting in a system of $N$ ordinary differential equations.
\end{remark}
%
%
For interpolation, we use a simple Gaussian kernel
\begin{equation*} 
    f_{e,i}(x, \gamma) = f_e(x_i, \gamma)
    \exp\left( - \frac{1}{2} \frac{\big( x-x_i(\gamma) \big)^2}{\sigma_i^2(\gamma)} \right) \enspace ,
\end{equation*}
\def\dxixj{x_i(\gamma)-x_j(\gamma)}
where $\sigma_i(\gamma)$ is chosen in such a way that the integral over $f_{e,i}(x, \gamma)$ is equal to one, which gives
\begin{equation*} 
    \sigma_i^2(\gamma) = \frac{2}{\pi} \big( \dxixj \big)^2 \enspace .
\end{equation*}
%
%
Finally, for a single sample, we obtain
\begin{equation} \label{Eq_NonParComponent}
    f_{e,i}(x, \gamma) 
    = \underbrace{\frac{1}{2} \frac{w_i}{ | x_i(\gamma) - x_j(\gamma) | }}_{\text{density value at sample}}
    \underbrace{\exp\left( - \frac{\pi}{4} \frac{\big( x-x_i(\gamma) \big)^2}
    {\big( \dxixj \big)^2} \right)}_{\text{interpolation between samples}} \enspace .
\end{equation}
%
%
For all samples, we have
\begin{equation*} 
    f_e(x, \gamma) = k(\gamma) \sum_{i=1}^L f_{e,i}(x, \gamma) \enspace ,
\end{equation*}
where $k(\gamma)$ is the factor that takes care of the fact that we calculate an unnormalized solution of the DODE \Eq{Eq_DODE_scalar_Gauss}.

%
%

\subsection{Derivation of DODE for Sample Locations}

%
%
We will now derive a distributed ordinary differential equation (DODE) that describes the evolution of the sample locations for the measurement update step along the artificial time $\gamma \in [0,1]$.
%
%
The derivative of $f_e(x, \gamma)$ with respect to $\gamma$ is
\begin{equation} \label{Eq_dotFe_NonPar}
    \begin{aligned}
        \dot{f}_e(x, \gamma)
        & = \dot{k}(\gamma) \sum_{i=1}^L f_{e,i}(x, \gamma) 
        + k(\gamma) \sum_{i=1}^L \dot{f}_{e,i}(x, \gamma) \\
        & = \frac{f_e(x, \gamma)}{k(\gamma)} \dot{k}(\gamma)
        + k(\gamma) \sum_{i=1}^L \dot{f}_{e,i}(x, \gamma) \enspace .
    \end{aligned}
\end{equation}
%
%
%
The derivatives of $f_{e,i}(x, \gamma)$ in \Eq{Eq_NonParComponent} with respect to $\gamma$ are given by
\def\dotetaxixj{\begin{bmatrix} \dot{x}_i(\gamma) \\ \dot{x}_j(\gamma) \end{bmatrix}}
\begin{equation*} 
    \dot{f}_{e,i}(x, \gamma) = \frac{\partial \, f_{e,i}(x, \gamma)}{\partial \, \gamma} 
    = \underbrace{\begin{bmatrix} 
        \frac{\pi \, x^2 - 2 \, x_i^2(\gamma)-x_j(\gamma) \big( \pi \, x + 2 \, x_j(\gamma) \big)
              + x_i(\gamma) \big( (4+\pi) x_j(\gamma) - \pi \, x \big)}{2 \, \big( \dxixj \big)^3} \\[4mm]
        -\frac{\pi \, x^2 - 2 \, \pi \, x \, x_i(\gamma) + \pi \, x_i^2(\gamma) 
        - 2 \big( \dxixj \big)^2}{2 \, \big( \dxixj \big)^3} 
    \end{bmatrix}^\top
    f_{e,i}(x, \gamma) }_{\vec{b}_i^\top(x,\gamma)}
    \dotetaxixj
\end{equation*}
for $i=1, \ldots, L$. We note that $\dot{f}_{e,i}(x, \gamma)$ only depends on the location $x_i(\gamma)$ with index $i$ (and its derivative $\dot{x}_i(\gamma))$) and on the location $x_j(\gamma)$ with index $j$, the nearest neighbor, (and its derivative $\dot{x}_j(\gamma))$.
The components of the vector $\vec{b}_i(x,\gamma)$ will be named as follows for the subsequent derivations
\begin{equation*}
    \vec{b}_i^\top(x,\gamma) = \begin{bmatrix} R_i(x,\gamma) & S_i(x,\gamma) \end{bmatrix} \enspace .
\end{equation*}
%
%
%
It can be rewritten as
\begin{equation*} 
    \vec{b}_i^\top(x,\gamma)
    = \begin{bmatrix} 1 & x & x^2 \end{bmatrix}
    \mat{A}_i(\gamma) \,
    f_{e,i}(x, \gamma)
\end{equation*}
with
\begin{equation*} 
    \mat{A}_i(\gamma) = \frac{1}{2 \, \big( \dxixj \big)^3}
    \small{%
    \begin{bmatrix} 
        (4+\pi) \, x_i(\gamma) \, x_j(\gamma) - 2 \, \big( x_i^2(\gamma) + x_j^2(\gamma) \big) 
        & 2 \, \big( \dxixj \big)^2 - \pi \, x_i^2(\gamma) \\[2mm]
        - \pi \, \big( x_i(\gamma) + x_j(\gamma) \big)
        & 2 \, \pi \, x_i(\gamma) \\[2mm]
        \pi & - \pi
    \end{bmatrix} \enspace .
    }%
\end{equation*}
%

%
%
The derivative of $\dot{f}_e(x, \gamma)$ in \Eq{Eq_dotFe_NonPar} can now be expressed as
\begin{equation*} 
    \dot{f}_e(x, \gamma) = \frac{f_e(x, \gamma)}{k(\gamma)} \dot{k}(\gamma)
    + k(\gamma) \sum_{i=1}^L 
    \begin{bmatrix} R_i(x,\gamma) & S_i(x,\gamma) \end{bmatrix}
    \dotetaxixj    
    \enspace .
\end{equation*}
which is equivalent to
\def\dotEtaVecSamples{
    \begin{bmatrix} 
        \dot{x}_1(\gamma) \\ 
        \dot{x}_2(\gamma) \\ 
        \vdots \\ 
        \dot{x}_L(\gamma) 
    \end{bmatrix}}
\begin{equation*} 
    \begin{alignedat}{4}
        \dot{f}_e(x, \gamma) 
        = \frac{f_e(x, \gamma)}{k(\gamma)} \dot{k}(\gamma)
        & + k(\gamma) \sum_{i=1}^L 
        \big[ 0 , 0 , \ldots , 0 , 
        \underbrace{R_i(x,\gamma)}_{\text{position } i} , 0 , \ldots , 0 , 0 \big]
        & \dotEtaVecSamples & \\
        & + k(\gamma) \sum_{i=1}^L 
        \big[ 0 , \ldots , 0 , 0 , 
        \underbrace{S_i(x,\gamma)}_{\text{position } j} , 0 , 0 , \ldots , 0 \big]
        & \dotEtaVecSamples &  
    \end{alignedat} \enspace .
\end{equation*}
The summations can be simplified to
\begin{equation*} 
    \begin{alignedat}{4}
        \dot{f}_e(x, \gamma) 
        = \frac{f_e(x, \gamma)}{k(\gamma)} \dot{k}(\gamma)
        & + k(\gamma) 
        \begin{bmatrix} R_1(x,\gamma), R_2(x,\gamma), \ldots, R_L(x,\gamma) \end{bmatrix}
        & \dotEtaVecSamples & \\
        & + k(\gamma)
        \begin{bmatrix} T_1(x,\gamma), T_2(x,\gamma), \ldots, T_L(x,\gamma) \end{bmatrix}
        & \dotEtaVecSamples &  
    \end{alignedat}
\end{equation*}
with
\begin{equation*}
    T_j(\gamma)=\sum_{\underset{j=\text{NN}(i)}{i=1}}^L S_i(\gamma)
\end{equation*}
for $j=1,\ldots,L$,
which leads to
\def\dotEtaVecSamplesK{
    \begin{bmatrix} 
        \dot{k}(\gamma) \\[2mm] 
        \dot{x}_1(\gamma) \\[2mm] 
        \dot{x}_2(\gamma) \\[2mm] 
        \vdots \\[2mm]
        \dot{x}_L(\gamma) 
    \end{bmatrix}}
\begin{equation} \label{Eq_rhsDODE_nonPar}
    \dot{f}_e(x, \gamma) = \mat{p}^\top(x, \gamma) \, \dot{\vec{\eta}}(\gamma)
\end{equation}
with
\begin{equation*}
    \mat{p}(x, \gamma) = 
    \begin{bmatrix} 
        \frac{f_e(x, \gamma)}{k(\gamma)} \\[2mm]
        k(\gamma) \big( R_1(x,\gamma) + T_1(x,\gamma) \big) \\[2mm]
        k(\gamma) \big( R_2(x,\gamma) + T_2(x,\gamma) \big) \\[2mm]
        \vdots \\[2mm]
        k(\gamma) \big( R_L(x,\gamma) + T_L(x,\gamma) \big)
    \end{bmatrix}
    \enspace ,
    \;\;\;
    \dot{\vec{\eta}}(\gamma) = \dotEtaVecSamplesK \enspace .
\end{equation*}

%
%
The DODE for the sample locations is now obtained by equating \Eq{Eq_rhsDODE_nonPar} with the left-hand side given in \Sec{Sec_Derivation_Zakai_DODE}.

%
%

\subsection{Solution of DODE for Sample Locations based on Collocation Points}

%
%
Solving this DODE is again performed by solving over a discrete set of points. 
Here, we focus on employing collocation points for an average solution over the state domain.
This leads to a system of coupled ordinary differential equations for the desired sample locations.
%
%
The collocation points are selected as follows. For each sample point $x_{e,i}(\gamma)$, a Dirac mixture approximation of the corresponding interpolation density $f_{e,i}\big(x, \vec{\eta}(\gamma) \big)$ is performed. These points $x_n$ for $n=1, \ldots, C$ are plugged into \Eq{Eq_rhsDODE_nonPar}, which gives 
\begin{equation*}
    \vec{q}\big( \vec{\eta}(\gamma) \big) 
    = \mat{P}\big( \vec{\eta}(\gamma) \big) \cdot \dot{\vec{\eta}}(\gamma) \enspace ,
\end{equation*}
with
\begin{equation*}
    \mat{P}\big( \vec{\eta}(\gamma) \big) 
    = \begin{bmatrix}  
        \mat{p}^\top(x_1, \gamma) \\[2mm]
        \mat{p}^\top(x_2, \gamma) \\[2mm]
        \vdots \\[2mm]
        \mat{p}^\top(x_C, \gamma)
    \end{bmatrix} \enspace .
\end{equation*}
%

%
%

\section{Conclusions} \label{Sec_Conclude}

%
%
The proposed new density approximation method generates a set of ordinary differential equations for the evolution of a parameter vector over an artificial time $\gamma \in [0,1]$ that starts with an approximation for the prior density for $\gamma=0$ and results in the desired approximation of the posterior for $\gamma=1$.
%
%
It provides some unique features compared with the current state of the art that make it well suited for practical applications:
\begin{itemize}
    \item The derivations are easy to understand.
    \item Implementation is straightforward and can be done based on standard ODE solvers.
    \item It is applicable to a wide variety of density approximation problems, where the approximated density does not even have to be known explicitly. This includes approximating the posterior densities in the Bayesian filter step in stochastic nonlinear dynamic systems.
    \item Solving for the desired posterior density parameters such as particle locations is inherently fast as neither an optimization nor solving a partial differential equation is required. 
\end{itemize}

%
%
Future work includes higher dimensions, varying the length of the parameter vector during the progression, and real-time implementations:
\begin{itemize}
    %
    %
    \item The derivations in this manuscript are limited on the scalar case for the sake of simplicity. The next step is the generalization to a higher number of dimensions.    
    %
    %
    \item The generated flow modifies the values of a parameter vector of fixed length in such a way that the posterior density is well approximated. 
    However, for both parametric and nonparametric representations, an adaptation of the number of parameters might be required. 
    For a particle representation, this means adding or removing particles depending on the local approximation quality.    
    %
    %
    \item So far, the algorithms are implemented in Matlab. For real-time operation, we envision to auto-generate C++ code that can be compiled for different platforms.
\end{itemize}

%
%

\bibliographystyle{StyleFiles/IEEEtran_Capitalize}

\bibliography{%
Literature/ISASPublikationen,%
Literature/ISASPublikationen_laufend,%
Literature/ISASPreprints,%
Literature/Literature%
}%

%
%

	    
    


\end{document}